\documentclass[letterpaper]{article} % DO NOT CHANGE THIS
\usepackage{aaai2027} % DO NOT CHANGE THIS
\nocopyright
\usepackage[hyphens]{url} % DO NOT CHANGE THIS
\usepackage{graphicx} % DO NOT CHANGE THIS
\usepackage{natbib} % DO NOT CHANGE THIS AND DO NOT ADD OPTIONS
\usepackage{caption} % DO NOT CHANGE THIS AND DO NOT ADD OPTIONS
\usepackage{amsmath,amssymb,amsfonts}
\usepackage{booktabs}
\usepackage{colortbl}
\definecolor{metricblue}{HTML}{B9D1E1}
\definecolor{metricorange}{HTML}{F2C38B}
\definecolor{metricgreen}{HTML}{C8DDC3}
\definecolor{metricgrayblue}{HTML}{C9D7DA}
\definecolor{ablationblue}{HTML}{7FA9C4}

\newcommand{\saBTCSharpe}{0.56}
\newcommand{\saBTCReturn}{7.02}
\newcommand{\saBTCMDD}{12.41}
\newcommand{\saETHSharpe}{0.23}
\newcommand{\saETHReturn}{1.17}
\newcommand{\saETHMDD}{16.45}

\title{SwingGuard: Risk-Conditioned Multi-Agent Swing Trading with Dual-Timescale Governance}

\author{
Junzhou Su\textsuperscript{\rm 1},
Yichen Luo\textsuperscript{\rm 2,3},
Yebo Feng\textsuperscript{\rm 4}\corresponding,
Teng Li\textsuperscript{\rm 1},
Jiahua Xu\textsuperscript{\rm 2}
}

\affiliations{
\textsuperscript{\rm 1}Xidian University\\
\textsuperscript{\rm 2}University College London\\
\textsuperscript{\rm 3}The University of Hong Kong\\
\textsuperscript{\rm 4}Nanyang Technological University\\
\texttt{yebo.feng@ntu.edu.sg}
}

\begin{document}
\maketitle

\begin{abstract}
Cryptocurrency swing trading spans mismatched timescales: positions remain
active while market and event risk can change within minutes.
However, many predictive and large language model (LLM)-based trading
systems assess risk after strategy generation or run both functions at the same
cadence, leaving proposals unconditioned by current risk and exposure weakly
governed between revisions. We propose SwingGuard, a dual-timescale, long-only multi-agent framework
for risk-conditioned strategy generation. At the
strategy timescale, specialized analysts synthesize point-in-time technical,
sentiment, on-chain, and macroeconomic evidence through bullish--bearish
deliberation. The resulting evidence and current risk context are then combined
by a Portfolio Architect to form a target position. At the governance
timescale, a five-minute Risk Controller performs conditional diagnosis,
maintains persistent risk states with staged recovery, audits the proposal, and
constrains execution until the next strategy revision. This design allows risk
to shape the feasible strategy before proposal finalization while continuing
to govern the active position afterward. Evaluations on Bitcoin (BTC) and
Ether (ETH) show that SwingGuard preserves positive returns after transaction costs
while maintaining shallow drawdowns and strong risk-adjusted performance on
both assets. The exposure-capped joint portfolio achieves a Sharpe ratio of
2.98, compared with 2.79 for an otherwise identical Post-hoc Risk variant.
Structured traces further link risk evidence and state transitions to strategy
constraints and executed position changes, providing an auditable framework
for asynchronous cryptocurrency swing trading. Overall, the results indicate
that conditioning strategy formation on current risk, together with persistent
high-frequency governance, improves risk-adjusted performance without
sacrificing auditability.
\end{abstract}

\section{Introduction}

Cryptocurrency swing trading creates an asynchronous control problem: a
strategy formed from a completed daily research window remains active while
prices, liquidity, on-chain activity, and event risk can change within
minutes. A medium-horizon thesis may therefore remain directionally valid
after its planned exposure becomes unsafe. Trading systems must coordinate
what position to seek with how much of that position remains executable and
when exposure may be restored~\cite{liu2021risks,liu2022common,
makarov2020trading,corbet2019crypto,fang2025cryptocurrency,
omrane2025volatility}.

Forecasting and reinforcement-learning methods primarily learn price or return
patterns and map predictive signals to trading decisions or portfolio
positions~\cite{jiang2016drl,llstm2019crypto,tharun2023deep,
cakici2024crypto,liu2023forecast,jing2024drl}. These representative methods
mainly optimize prediction or strategy decisions without explicitly modeling
the independent fast risk-governance layer studied here. Sequence forecasters
model long-horizon and general temporal structure~\cite{zhou2021informer,wu2021autoformer,
nie2023patchtst,wu2023timesnet}. Large language models (LLMs) support increasingly structured forms of
reasoning and financial-domain adaptation~\cite{wei2022cot,
xiong2025deliberate,xiong2026enhancing,liu2023fingpt,xie2023pixiu}, while LLM-based multi-agent trading systems use
complementary mechanisms including specialized analysis, agent coordination,
deliberation, and memory~\cite{luo2025llmagents,
yu2024fincon,xiao2024tradingagents}. Yet proposal-only risk review neither shapes the
preceding deliberation nor governs the position at a faster cadence.

The central distinction is intervention timing. For strategy information
$X_t$, policy $\pi$, and risk context $R_t$, a post-hoc operator
$g(\pi(X_t),R_t)$ modifies a completed proposal, whereas $\pi(X_t,R_t)$ lets
risk alter direction, exposure, and parameters during deliberation. We ask
whether this conditioning improves the return--risk trade-off over an
otherwise identical post-hoc design while preserving an auditable path to
execution.

We propose \textbf{SwingGuard}, a dual-timescale multi-agent architecture
for risk-conditioned cryptocurrency swing trading. Technical, sentiment,
and fundamentals analyzers inform bullish--bearish deliberation, after which
a Portfolio Architect forms a long-only target. A coupled five-minute risk
chain diagnoses threats, maintains a persistent state with staged recovery,
and constrains execution. Its current risk context enters the Portfolio
Architect before proposal finalization; a deterministic audit subsequently
enforces the controller boundaries. The strategy and governance layers operate
at distinct timescales: strategy reasoning occurs over the completed research
window, while risk governance updates every five minutes.

We evaluate Bitcoin (BTC) and Ether (ETH) from August 2024 through January
2025 against passive,
rule-based, machine-learning, and single-agent references using cost-adjusted
performance and trace-level evidence.

The main contributions of this paper are summarized as follows:

\begin{itemize}
\item \textbf{Risk-conditioned strategy generation:} the current System Risk Context constrains the feasible strategy before the Portfolio Architect finalizes a proposal.

\item \textbf{Dual-timescale risk governance:} daily structural reasoning is coordinated with five-minute detection, persistent state transitions, staged recovery, and execution constraints.

\item \textbf{Auditable risk-to-execution linkage:} structured records connect risk evidence and state transitions to strategy constraints and executed positions.
\end{itemize}

\section{Related Work}

Cryptocurrency forecasting and reinforcement learning estimate returns or map
signals to positions~\cite{zhang2024deepcrypto,john2024survey,jiang2016drl,
liu2023forecast,jing2024drl}, while technical, news, and momentum evidence
remains regime sensitive~\cite{hudson2021technical,anamika2022news,
asness2013momentum,cakici2024crypto}. Volatility-managed and regime-switching
methods instead treat exposure and persistent market state as control
objects~\cite{moreira2017volatility,harvey2018volatility,hamilton1989regime}.
SwingGuard builds on this distinction between a directional thesis and executable
exposure.

Financial LLMs combine domain resources~\cite{li2023llmfinance,liu2023fingpt,
xie2023pixiu} with trading-oriented reflection and specialized reasoning
~\cite{koa2024reflection,li2024cryptoagent,wang2024factreasoning,
ding2024tradexpert}. Multi-agent frameworks add structured roles
~\cite{wu2023autogen,hong2023metagpt}, debate~\cite{du2023debate,
liang2023debate}, and scalable collaboration~\cite{qian2025scaling}.
Beyond role specialization, recent work on general LLM reasoning increasingly
introduces explicit structure and control into long-horizon inference.
Structure-aware planning organizes intermediate reasoning around structured
state transitions and verification~\cite{xiong2025deliberate}, multi-level
reasoning decomposes long trajectories into high-level planning and low-level
execution~\cite{xiong2026enhancing}, and adaptive information control regulates
the extent and granularity of evidence acquisition in search-augmented
reasoning~\cite{xiong2026adaptive}. These approaches motivate structured
reasoning and controlled information flow, but address general reasoning or
retrieval rather than financial risk governance. Financial systems apply
role-based and deliberative mechanisms to research and trading
~\cite{yu2024fincon,xiao2024tradingagents}, while recent cryptocurrency
applications employ multi-agent reasoning in portfolio allocation and
adversarial trading environments~\cite{luo2025llmagents,luo2026resisting}.
SwingGuard instead addresses the asynchronous case: faster risk updates condition
strategy generation and govern the active swing position. Beyond post-hoc
attribution and the broader interpretable-ML literature
~\cite{lundberg2017shap,rudin2019interpretable,yeo2023comprehensive}, it links
evidence, risk states, constraints, and realized positions through an auditable
strategy--risk interface.

\section{Methodology}
\label{sec:methodology}

We first formulate the asynchronous decision problem, then describe the
strategy and risk chains and their mapping to executed exposure.

\subsection{Problem Formulation}

Let $d$ index a completed daily strategy-generation window and $\tau$ a
subsequent five-minute governance instant. The strategy input
$X_d=(M_d,N_d,F_d)$ contains technical market, news and event sentiment,
and on-chain and macroeconomic evidence observable by the close of $d$.
Let $Z_{d,\tau}$ be the latest completed five-minute observation. From it,
the controller assigns $S_{d,\tau}$ to NORMAL, ALERT, SURVIVAL, or RECOVERY
and exposes the \emph{System Risk Context}
\begin{equation}
R_{d,\tau}=(S_{d,\tau},q_{d,\tau},\bar w_{d,\tau},
\xi_{d,\tau},\ell_d),
\label{eq:risk_context}
\end{equation}
where $q$ is the composite risk score, $\bar w$ the position cap, $\xi$ supporting
short-horizon evidence, and $\ell$ the long-horizon market regime.

Let $D_d$ denote analysis and deliberation, and $U_d$ the carried position
and strategy feedback. The Portfolio Architect produces a raw long-only
target $\widehat w_d$ and executable parameters $\Theta_d$:
\begin{equation}
\begin{aligned}
(\widehat w_d^{\mathrm{full}},\Theta_d^{\mathrm{full}})
  &=\pi(D_d,R_{d,0},U_d),\\
(\widehat w_d^{\mathrm{post}},\Theta_d^{\mathrm{post}})
  &=\pi(D_d,U_d).
\end{aligned}
\label{eq:full_proposal}
\end{equation}
Let $p(d,\tau)$ denote the preceding execution instant and
$w^{\mathrm{exec}}_{p(d,\tau)}$ its realized position. Both variants then use
the same deterministic audit-and-arbitration mapping:
\begin{equation}
w_{d,\tau}=g(\widehat w_d,\Theta_d,R_{d,\tau},
w^{\mathrm{exec}}_{p(d,\tau)}),\qquad
0\leq w_{d,\tau}\leq\bar w_{d,\tau}.
\label{eq:risk_constraint}
\end{equation}
Here $g$ abbreviates the two-stage deterministic application of audit $h$
followed by arbitration $a$, detailed below. Thus the ablation changes when risk enters reasoning, not the
downstream safety mechanism. Targets formed from completed observations
apply only to the next return interval.

The distinction can be stated as a difference in feasible strategy spaces.
Let $\Theta_d$ collect entry, exit, stop, validity-horizon, and sizing
parameters, let $\Gamma_d=(\widehat w_d,\Theta_d)$ denote the complete
proposal, and let $\mathcal F(R_{d,0})$ denote the set admitted by the current
risk state. Risk-conditioned generation and post-hoc control
can be represented as
\begin{equation}
\begin{aligned}
\Gamma_d^{\mathrm{full}}
  &=\pi(D_d,U_d;\mathcal F(R_{d,0})),\\
\Gamma_d^{\mathrm{post}}
  &=\mathcal P_{\mathcal F(R_{d,0})}
    [\pi(D_d,U_d;\mathcal F_0)],
\end{aligned}
\label{eq:conditioning_projection}
\end{equation}
where $\mathcal F_0$ is the risk-agnostic strategy space under the fixed
long-only rules and
$\mathcal P$ is the downstream audit projection. These operators are not
generally equivalent: projection can enforce safety bounds, but cannot
reconstruct the rationale and coupled parameters that would have been formed
inside the risk-conditioned feasible set.

\subsection{Dual-Timescale SwingGuard Framework}

\begin{figure*}[t]
\centering
\includegraphics[width=0.9\textwidth]{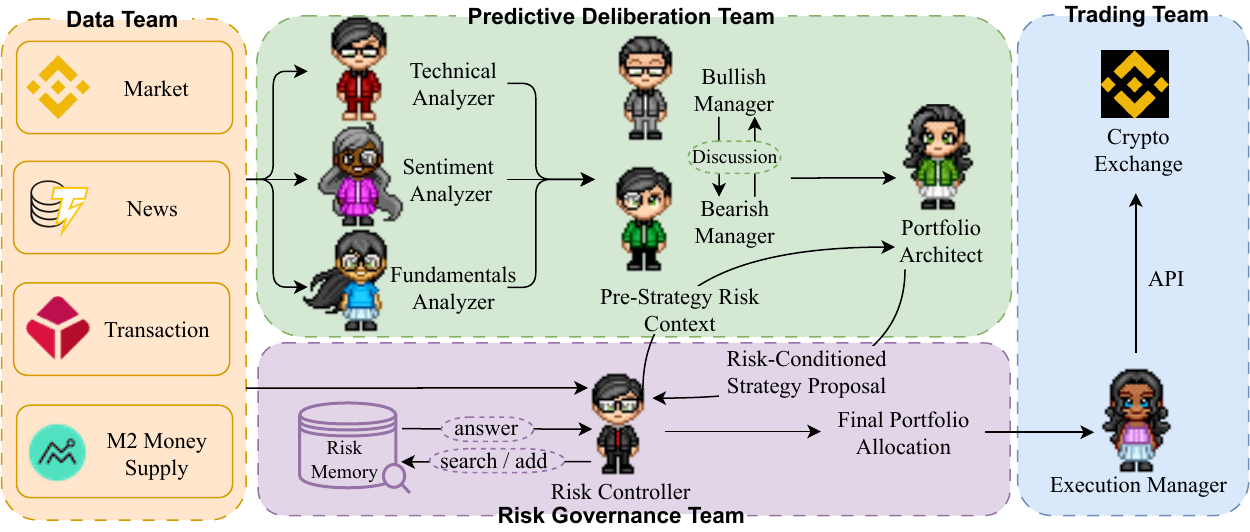}
\caption{Overall SwingGuard architecture. The Data Team provides market, news, and
on-chain transaction evidence to the analytical agents and adds the latest
observable U.S. M2 broad-money context before deliberation. The Risk Controller supplies pre-generation risk
context to the Portfolio Architect, audits the proposed target position, and
continues five-minute governance before the validated target reaches
the Execution Manager.}
\label{fig:architecture}
\end{figure*}

Figure~\ref{fig:architecture} shows two coupled loops. The deterministic Data
Team aligns point-in-time sources for the daily analytical loop, which combines
multimodal evidence and bullish--bearish debate into $\widehat w_d$. The
five-minute loop supplies $R_{d,0}$ before generation, audits the proposal, and updates the
executable cap until the next daily cycle without rerunning the strategy LLMs.
The deterministic Execution Manager converts approved position changes into
exchange actions.
This manager--specialist organization follows established role-based agent
coordination, while the explicit controller boundary restricts open-ended
conversation with a deterministic execution interface
~\cite{wu2023autogen,hong2023metagpt,qian2025scaling}.

\subsection{Daily Multi-Agent Strategy Generation}

The Technical Analyzer, Sentiment Analyzer, and Fundamentals Analyzer
transform $X_d$ into technical-market, news-event-sentiment, and
on-chain-macroeconomic reports,
\begin{equation}
A_d=(\phi_M(M_d),\phi_N(N_d),\phi_F(F_d)).
\label{eq:analyst_reports}
\end{equation}
The Bullish and Bearish Managers receive the same reports and construct
opposing long-only interpretations: the former argues for exposure, while the
latter argues for reduction or cash rather than short selling. This deliberation
exposes competing reasoning paths before synthesis
~\cite{du2023debate,liang2023debate}. Their trace is
\begin{equation}
B_d=\beta(A_d,U_d),\qquad D_d=(A_d,B_d).
\label{eq:bull_bear_debate}
\end{equation}
The Portfolio Architect internally uses a Research Manager to synthesize
$D_d$ and a Trader to convert that plan into a target and executable
parameters $\Theta_d$ (entry, stop, take-profit, validity, and size).
Full SwingGuard supplies $R_{d,0}$ to both internal stages, so the risk state can alter the direction,
feasible size, and parameters considered before finalization. Post-hoc
Risk omits this input while holding reports, debate, strategy memory, and
downstream control fixed.

\subsection{Five-Minute Hierarchical Risk Diagnosis}

\begin{figure*}[t]
\centering
\begin{tabular}{@{}cc@{}}
\includegraphics[width=0.446\textwidth]{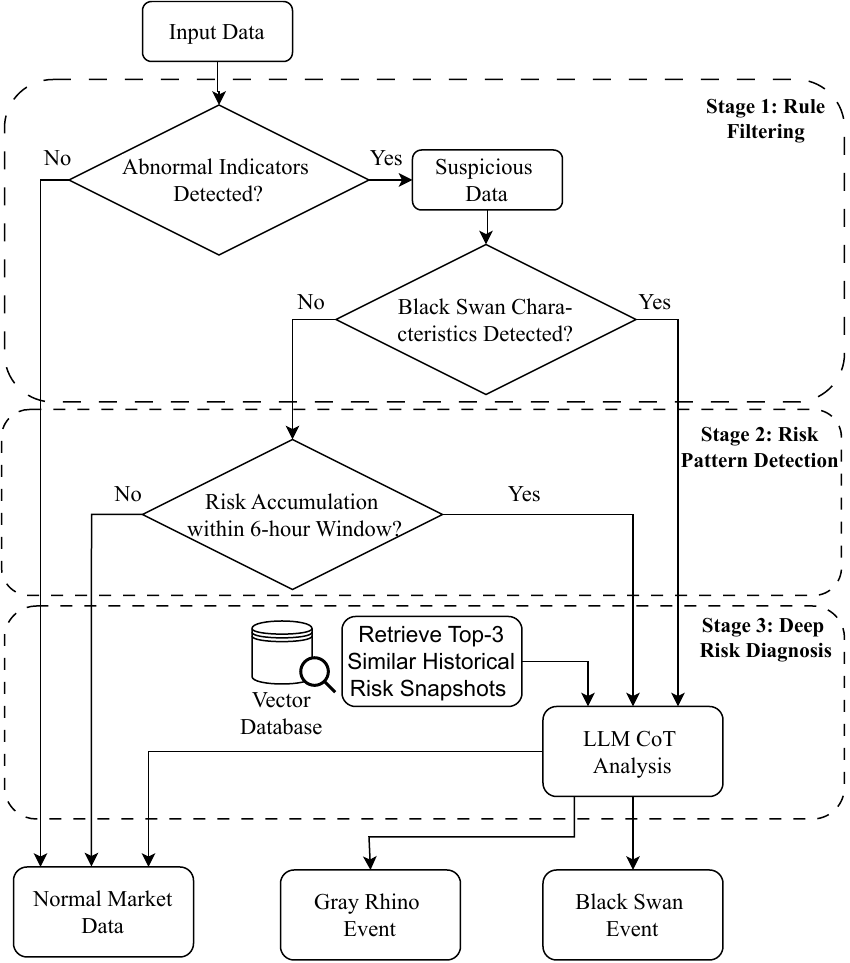} &
\includegraphics[width=0.382\textwidth]{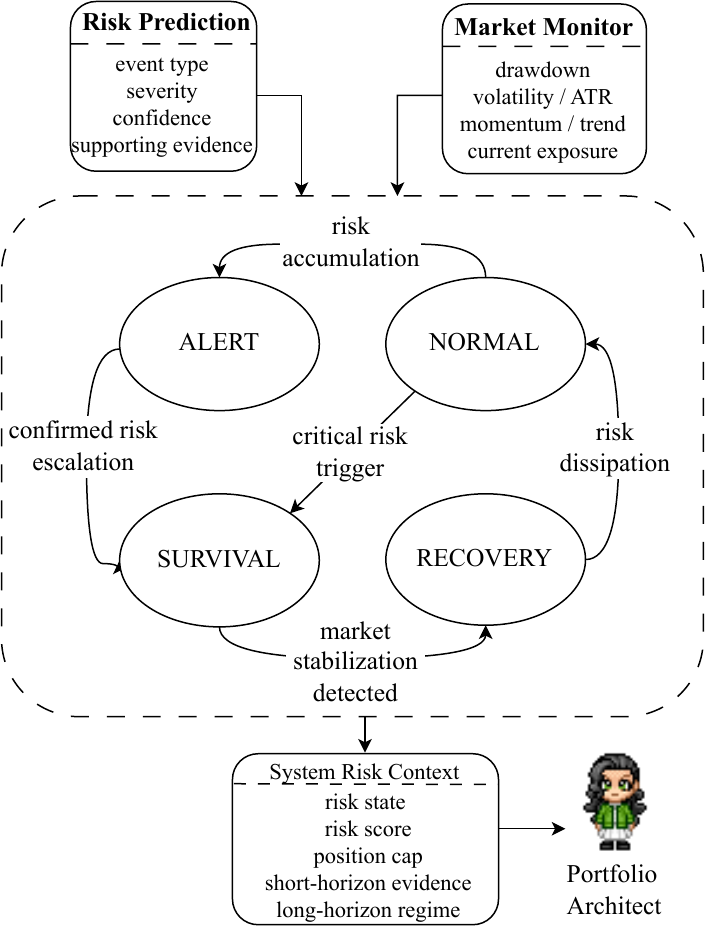} \\
\textbf{(a) Three-Stage Risk Detection and Diagnosis.} &
\textbf{(b) Persistent Risk-State Transition and Strategy Conditioning.}
\end{tabular}
\caption{Hierarchical risk diagnosis and control framework. Panel (a) uses
rule screening and six-hour persistence to trigger retrieval-augmented
diagnosis. Panel (b) combines confirmed evidence and market conditions into
persistent state transitions, producing the System Risk Context for strategy
generation. ``LLM CoT'' records Stage~3's bounded diagnostic rationale.}
\label{fig:risk_radar}
\end{figure*}

As summarized in Figure~\ref{fig:risk_radar}, for every completed five-minute
bar, Stage~1 screens market, on-chain, and news anomalies with deterministic
rules. Stage~2 evaluates multi-source
accumulation, recent density, and source diversity over a six-hour buffer.
Only an accumulated trigger $\gamma_{d,\tau}$ or designated severe event
$b_{d,\tau}$ invokes Stage~3:
\begin{equation}
\mathcal{D}_{d,\tau}=
\begin{cases}
\mathcal{L}(Z_{d,\tau},\mathcal{C}_{d,\tau},\mathcal{K}_{d,\tau}),
& \gamma_{d,\tau}\vee b_{d,\tau},\\
\varnothing,&\text{otherwise},
\end{cases}
\label{eq:conditional_diagnosis}
\end{equation}
where $\mathcal{C}_{d,\tau}$ is recent context and
$\mathcal{K}_{d,\tau}$ historical context. The LLM classifies the current
condition as normal, gray-rhino risk, or black-swan risk, with confidence and
supporting evidence; a deterministic fallback handles invalid outputs. Conditional
invocation concentrates LLM reasoning on persistent or severe events.

For historical context, each archived snapshot $e_i$ stores symbol $s_i$,
metadata time $t_i^{\mathrm{meta}}$, diagnostic fields, and a vector
$\boldsymbol v_i$. A fixed, normalized 14-dimensional vector summarizes market,
transaction, and news conditions. For traded symbol $s_d$, the current vector
$\boldsymbol v_{d,\tau}$ retrieves only earlier same-symbol snapshots:
\begin{equation}
\mathcal{K}_{d,\tau}
=\left\{e_i:i\in\underset{
i:\,s_i=s_d,\;t_i^{\mathrm{meta}}<t_{d,\tau}}{\operatorname{Top3}}
\cos(\boldsymbol v_{d,\tau},\boldsymbol v_i)\right\}.
\label{eq:risk_memory_retrieval}
\end{equation}
Here, $\operatorname{Top3}$ returns the indices of the three highest cosine
similarities. The retrieved snapshots provide precursors, causes, drawdown, and
recovery profiles for diagnosis. This design draws on retrieval augmentation
and episodic-memory mechanisms~\cite{lewis2020rag,shinn2023reflexion}, while
keeping the historical archive distinct from the online persistence state:
the former supplies similar-event evidence for diagnosis, whereas the latter
tracks whether current risk is accumulating, dissipating, or recovering.

\subsection{Persistent Risk State and Recovery}

The \emph{Risk Prediction} interface normalizes event type, severity,
confidence, and supporting evidence, while the \emph{Market Monitor} supplies
drawdown, volatility/average true range (ATR), momentum/trend, and current
exposure. Their fusion produces $q_{d,\tau}$. Given predecessor $p(d,\tau)$,
the state evolves
with active events $\mathcal E$, hold counters $\boldsymbol u$, and recovery
evidence $\rho$:
\begin{equation}
S_{d,\tau}=T(S_{p(d,\tau)},q_{d,\tau},
\mathcal E_{d,\tau},\boldsymbol u_{d,\tau},\rho_{d,\tau}).
\label{eq:risk_transition}
\end{equation}
Hysteresis and minimum holds prevent a quiet bar from immediately erasing
risk. Recovery requires jointly satisfied hold, price, volatility, quiet, and
market-repair predicates:
\begin{equation}
\rho_{d,\tau}=\mathbb I[c^{\mathrm{hold}}\wedge c^{\mathrm{price}}
\wedge c^{\mathrm{vol}}\wedge c^{\mathrm{quiet}}
\wedge c^{\mathrm{repair}}].
\label{eq:recovery_evidence}
\end{equation}
NORMAL permits the standard cap; ALERT compresses exposure; SURVIVAL
forces zero; and RECOVERY restores exposure in stages. Renewed risk can
re-escalate any recovery. The resulting control is
\begin{equation}
(\bar w_{d,\tau},a^{\mathrm{risk}}_{d,\tau})
=\psi(S_{d,\tau},s^{\mathrm{rec}}_{d,\tau},
\mathcal E_{d,\tau},E^{\mathrm{int}}_{d,\tau}).
\label{eq:state_position_cap}
\end{equation}
Here $s^{\mathrm{rec}}$ is the staged-recovery level,
$E^{\mathrm{int}}$ is the active set of intervention constraints, and
$a^{\mathrm{risk}}$ is the resulting controller action.
At $\tau=0$, the controller first consumes the latest completed bar,
updates the state, and serializes Eq.~\eqref{eq:risk_context} before
building either strategy prompt. This ordering implements pre-strategy
conditioning without seeing the subsequent return.

\subsection{Risk-Conditioned Strategy Generation, Audit, and Execution}

After proposal generation, a hard-boundary auditor $h$ validates long-only
direction, size, stop, validity horizon, and controller limits; an arbiter
$a$ applies the current state action:
\begin{align}
(\widetilde w_{d,\tau},V_{d,\tau})
&=h(\widehat w_d,\Theta_d,R_{d,\tau}),
\label{eq:hard_boundary_audit}\\
w_{d,\tau}
&=a(\widetilde w_{d,\tau},R_{d,\tau},w^{\mathrm{exec}}_{p(d,\tau)}).
\label{eq:risk_arbiter_mapping}
\end{align}
The audit verdict $V_{d,\tau}$ records any validation or clipping decision.
Until the next five-minute open, the preceding executed position remains
active; at that open, $w^{\mathrm{exec}}_{d,\tau}=w_{d,\tau}$ and
$a^{\mathrm{exec}}_{d,\tau}$ records the corresponding position change.
Both Full and Post-hoc Risk retain this audit; only Eq.~\eqref{eq:full_proposal}
differs. Between daily updates, $R_{d,\tau}$ may reduce, liquidate, or stage
re-entry without regenerating the strategy. The Execution Manager maps the
change from the preceding executed position to OPEN, ADD, HOLD, REDUCE, or
CLOSE and records
\begin{equation}
\begin{aligned}
\mathcal Q_{d,\tau}=(&R_{d,\tau},\Theta_d,\widehat w_d,V_{d,\tau},\\
&\widetilde w_{d,\tau},w_{d,\tau},a^{\mathrm{exec}}_{d,\tau},
w^{\mathrm{exec}}_{d,\tau}).
\end{aligned}
\label{eq:audit_record}
\end{equation}
This trace separates the LLM proposal, validated target, state-driven
constraint, and realized position, linking risk evidence to execution.

\section{Experimental Setup}
\label{sec:experimental_setup}

We evaluate cost-adjusted value, cross-asset consistency, and
risk-to-execution auditability. External baselines measure comparative
performance; controlled variants test architectural effects.

\paragraph{Data and protocol.}
We evaluate long-only BTC and ETH from August 1, 2024 through January 31,
2025. Each asset provides 52,992 five-minute observations; earlier data
support indicator warm-up and Random Forest training. Decisions use only
completed bars, preceding-seven-day news, observable on-chain records, and
the latest released U.S. M2 broad-money observation (series code WM2NS);
ex-post regimes
serve analysis only.
Single-asset tests start with unit wealth and zero exposure. The joint
portfolio synchronizes executed BTC and ETH positions without another LLM
decision. For $w_t^B,w_t^E\in[0,1]$,
\begin{equation}
(\omega^B_t,\omega^E_t)=
\begin{cases}
(w^B_t,w^E_t), & w^B_t+w^E_t\leq 1,\\[2pt]
\dfrac{(w^B_t,w^E_t)}{w^B_t+w^E_t},
& w^B_t+w^E_t>1.
\end{cases}
\label{eq:portfolio_scaling}
\end{equation}
Gross joint return is $\omega_t^B r_{t+1}^B+\omega_t^E r_{t+1}^E$; costs
are charged on both weight changes after scaling.

\paragraph{Models and baselines.}
The controlled SwingGuard variants share the GPT-4o backbone,
zero-temperature decoding, a 2,048-token response limit, identical prompts for
retained components, and all other non-ablated settings; variants retaining the
debate module use one round.
Baselines include buy-and-hold and a 20-day moving average (MA); a 4-hour
Donchian--ATR rule motivated by prior cryptocurrency technical-trading
evidence~\cite{hudson2021technical} and the average true range
(ATR)~\cite{wilder1978new}; a 4-hour ATR trend-exposure rule using momentum
signals~\cite{asness2013momentum}; a pre-evaluation 4-hour Random Forest
strategy~\cite{breiman2001random,gu2020ml}; a daily volatility-managed rule
that scales exposure toward 35\% annualized volatility estimated over 20
days~\cite{moreira2017volatility,harvey2018volatility};
an observable regime-switching rule combining a 120-day trend with 20-day
volatility relative to its trailing 120-day median~\cite{hamilton1989regime};
and a Daily Single-Agent LLM. The
latter uses the same model, decoding settings, token limit, point-in-time
modalities, and daily cadence, but a single tool-free prompt that directly
outputs a long-only target exposure in $[0,1]$, held until the next revision.

\paragraph{Controlled comparisons.}
Table~\ref{tab:ablation_matrix} supports two adjacent risk comparisons and
two supporting component removals. Full versus Post-hoc Risk changes only
whether $R_{d,0}$ enters strategy generation while retaining the same
subsequent audit and controller. Post-hoc Risk versus No Risk Controller then
changes only the downstream audit and five-minute governance while keeping
pre-generation conditioning absent; the strategy-native stop remains active.
The final two variants remove the three analyst reports as one group or
remove bullish--bearish deliberation. Point-in-time source windows and unaffected
upstream outputs are held fixed, while affected downstream decisions are
regenerated.
\begin{table}[!htbp]
\centering
\small
\setlength{\tabcolsep}{3pt}
\begin{tabular*}{\columnwidth}{@{\extracolsep{\fill}}lcccc@{}}
\toprule
Mode & Pre & Post & Analysts & Debate \\
\midrule
Full SwingGuard & \cellcolor{metricblue!22}Yes & \cellcolor{metricblue!22}Yes & \cellcolor{metricblue!22}Yes & \cellcolor{metricblue!22}Yes \\
Post-hoc Risk & No & \cellcolor{metricblue!22}Yes & \cellcolor{metricblue!22}Yes & \cellcolor{metricblue!22}Yes \\
No Risk Controller & No & No & \cellcolor{metricblue!22}Yes & \cellcolor{metricblue!22}Yes \\
No Analyst Team & \cellcolor{metricblue!22}Yes & \cellcolor{metricblue!22}Yes & No & \cellcolor{metricblue!22}Yes \\
No Bull/Bear Debate & \cellcolor{metricblue!22}Yes & \cellcolor{metricblue!22}Yes & \cellcolor{metricblue!22}Yes & No \\
\bottomrule
\end{tabular*}
\caption{Controlled SwingGuard configurations. Pre and Post denote risk
conditioning before strategy generation and risk governance afterward.}
\label{tab:ablation_matrix}
\end{table}

\paragraph{Execution and metrics.}
Signals execute at the next five-minute open. Every method uses the same
prices, long-only constraint, 10-basis-point commission, and 5-basis-point
slippage, for a 15-basis-point one-way cost; turnover is charged whenever
executed exposure changes. Primary comparisons
use cumulative net return, annualized Sharpe~\cite{sharpe1994ratio}, and
maximum drawdown. We additionally compute annualized volatility,
Sortino~\cite{sortino1994performance}, Calmar, turnover,
and average exposure from the five-minute marked-to-market series, using a
365-day cryptocurrency calendar and zero
risk-free rate. Monthly net returns
measure temporal concentration, while single-asset and exposure-capped
portfolio results measure cross-asset consistency.

\paragraph{Look-ahead bias control.}

All decision inputs obey a point-in-time availability boundary. For a
decision timestamp $t$, technical features use only bars closed by $t$;
the news window ends at $t$; on-chain observations must already be
observable; and WM2NS contributes only the latest observation whose release
date has occurred by $t$. The Random Forest is trained exclusively on data
before August 1, 2024 and is frozen throughout evaluation. SwingGuard and every
baseline form signals from the same historical boundary, and any resulting
position change is executed no earlier than the next five-minute open.
Finally, ex-post regime labels and future local troughs are used only for
retrospective analysis and figure annotation; neither enters agent prompts,
baseline features, portfolio construction, nor ablation regeneration.

\section{Experimental Results}
\label{sec:results}

\subsection{Cross-Asset Economic Performance}

Table~\ref{tab:main_results} compares cost-adjusted BTC and ETH performance.
SwingGuard is the only reported method that produces double-digit positive
returns on both assets while maintaining Sharpe ratios above 2 and maximum
drawdown (MDD) below 6\%. It also has the highest Sharpe and smallest drawdown
point estimates, together with the lowest reported annualized volatility, in both
single-asset comparisons.

\begin{table*}[t]
\centering
\small
\setlength{\tabcolsep}{3.2pt}
\begin{tabular*}{\textwidth}{@{\extracolsep{\fill}}lcccccccc@{}}
\toprule
& \multicolumn{4}{c}{BTC} & \multicolumn{4}{c}{ETH} \\
\cmidrule(lr){2-5}\cmidrule(lr){6-9}
Method & Sharpe $\uparrow$ & Return (\%) $\uparrow$ & MDD (\%) $\uparrow$ & Vol. (\%) $\downarrow$ &
Sharpe $\uparrow$ & Return (\%) $\uparrow$ & MDD (\%) $\uparrow$ & Vol. (\%) $\downarrow$ \\
\midrule
Buy-and-hold & \cellcolor{metricblue!25}1.94 & \cellcolor{metricorange!60}\textbf{58.25} & -24.70 & 54.45 &
\cellcolor{metricblue!25}0.40 & \cellcolor{metricorange!25}1.96 & -33.06 & 68.97 \\
MA (20d) & 1.91 & \cellcolor{metricorange!25}41.69 & -15.32 & 40.52 & -0.52 & -14.53 & -31.10 & 42.34 \\
Donchian--ATR 4h & \cellcolor{metricblue!40}1.95 & 36.94 & \cellcolor{metricgreen!25}-12.22 & \cellcolor{metricgrayblue!25}35.22 &
0.28 & 1.78 & -23.64 & 38.58 \\
ATR trend exposure 4h & 1.34 & 17.29 & \cellcolor{metricgreen!40}-10.91 & \cellcolor{metricgrayblue!40}26.25 &
0.07 & -0.67 & \cellcolor{metricgreen!25}-16.74 & \cellcolor{metricgrayblue!40}25.06 \\
Random Forest 4h & 0.50 & 6.00 & -21.98 & 36.75 & -0.54 & -12.94 & -28.76 & 37.91 \\
Volatility managed & 1.93 & \cellcolor{metricorange!40}42.13 & -19.70 & 40.23 & 0.22 & 0.60 & -22.31 & 38.19 \\
Observable regime switching & 1.52 & 27.54 & -18.50 & 36.00 &
\cellcolor{metricblue!40}0.68 & \cellcolor{metricorange!40}9.23 & -22.48 & \cellcolor{metricgrayblue!25}34.75 \\
Daily Single-Agent LLM & \saBTCSharpe & \saBTCReturn & -\saBTCMDD & 36.79 &
\saETHSharpe & \saETHReturn & \cellcolor{metricgreen!40}-\saETHMDD & 35.64 \\
Full SwingGuard & \cellcolor{metricblue!60}\textbf{2.70} & 19.91 & \cellcolor{metricgreen!60}\textbf{-3.74} & \cellcolor{metricgrayblue!60}\textbf{13.67} &
\cellcolor{metricblue!60}\textbf{2.36} & \cellcolor{metricorange!60}\textbf{15.35} & \cellcolor{metricgreen!60}\textbf{-5.78} & \cellcolor{metricgrayblue!60}\textbf{12.32} \\
\bottomrule
\end{tabular*}
\caption{Cost-adjusted performance from August 2024 to January 2025. MDD is
reported as a negative return; Vol. denotes annualized volatility. Arrows show
the preferred direction. Blue, orange, green, and gray-blue identify the top
three results per column for Sharpe, return, MDD, and volatility, respectively;
the best result is boldfaced.}
\label{tab:main_results}
\end{table*}

Volatility management preserves much of the BTC opportunity (42.13\% return
and 1.93 Sharpe) but transfers weakly to ETH. Observable regime switching
improves ETH return to 9.23\% and Sharpe to 0.68, yet retains 22.48\% drawdown.
SwingGuard remains profitable on both assets, exceeds passive ETH return by 13.39
percentage points, and reduces ETH drawdown by 82.5\%. Its higher Sharpe and
lower drawdown on both assets are not reproduced by either simple exposure
scaling or an observable trend--volatility filter. Its BTC and ETH annualized
volatilities, 13.67\% and 12.32\%, are approximately half the next-lowest
reported values of 26.25\% and 25.06\%, respectively.

The Daily Single-Agent LLM aligns the daily cadence and continuous long-only
position space, providing a direct single-agent reference. It is profitable on
both assets but transfers weakly to ETH, whereas SwingGuard preserves double-digit
ETH return and improves BTC Sharpe and drawdown. The remaining system-level
distinction is SwingGuard's specialized roles, deliberation, and five-minute risk
governance.

\subsection{Capital Efficiency and Temporal Concentration}

SwingGuard attains this risk-adjusted performance with limited capital deployment:
average exposure is 19.42\%
for BTC, 12.56\% for ETH, and 31.74\% after the joint cap. The synchronized,
cost-adjusted portfolio returns 37.16\% with 2.98 Sharpe and 6.44\% drawdown,
versus 30.11\%, 1.19, and 28.47\% for equal-weight buy-and-hold. Because the
joint result uses capped positions rather than summed returns, its comparison
with buy-and-hold reflects active timing, cash allocation, proportional
exposure scaling, and associated trading costs under a shared capital budget.
BTC, ETH, and joint turnover are 14.54, 11.50, and 25.80.
Repricing the same executed positions at 5, 30, and 50 basis points one-way
cost leaves joint returns of 40.75\%, 31.95\%, and 25.31\% and Sharpe ratios
of 3.21, 2.62, and 2.15, respectively.

With 31.74\% average exposure, the capped portfolio exceeds equal-weight
buy-and-hold return by 7.05 percentage points, raises Sharpe from 1.19 to 2.98,
and reduces drawdown magnitude by 77.4\%, linking selective deployment to
capital efficiency and downside control.

Downside-sensitive metrics reinforce the shallow-drawdown result. BTC and ETH
obtain Sortino ratios of 3.89 and 3.36 and Calmar ratios of 11.60 and 5.66;
the capped portfolio reaches 4.26 and 13.54, respectively. Sortino ratios above
the corresponding Sharpe ratios indicate that downside variation is smaller
than total return variation. Separately, low average exposure shows that this
risk-adjusted performance is achieved through selective capital deployment
rather than continuous market participation.

BTC, ETH, and the portfolio are profitable in five of six months. November
provides the largest common opportunity, while divergent December and January
exposures offset losses across assets. Returns are therefore concentrated but
not confined to a single month: the largest single-asset monthly loss is
0.94\%, and the portfolio's sole monthly loss is 0.02\%. SwingGuard captures the
principal opportunity while preserving gains as the assets later weaken or
diverge.

\begin{figure}[!htbp]
\centering
\includegraphics[width=0.87\columnwidth]{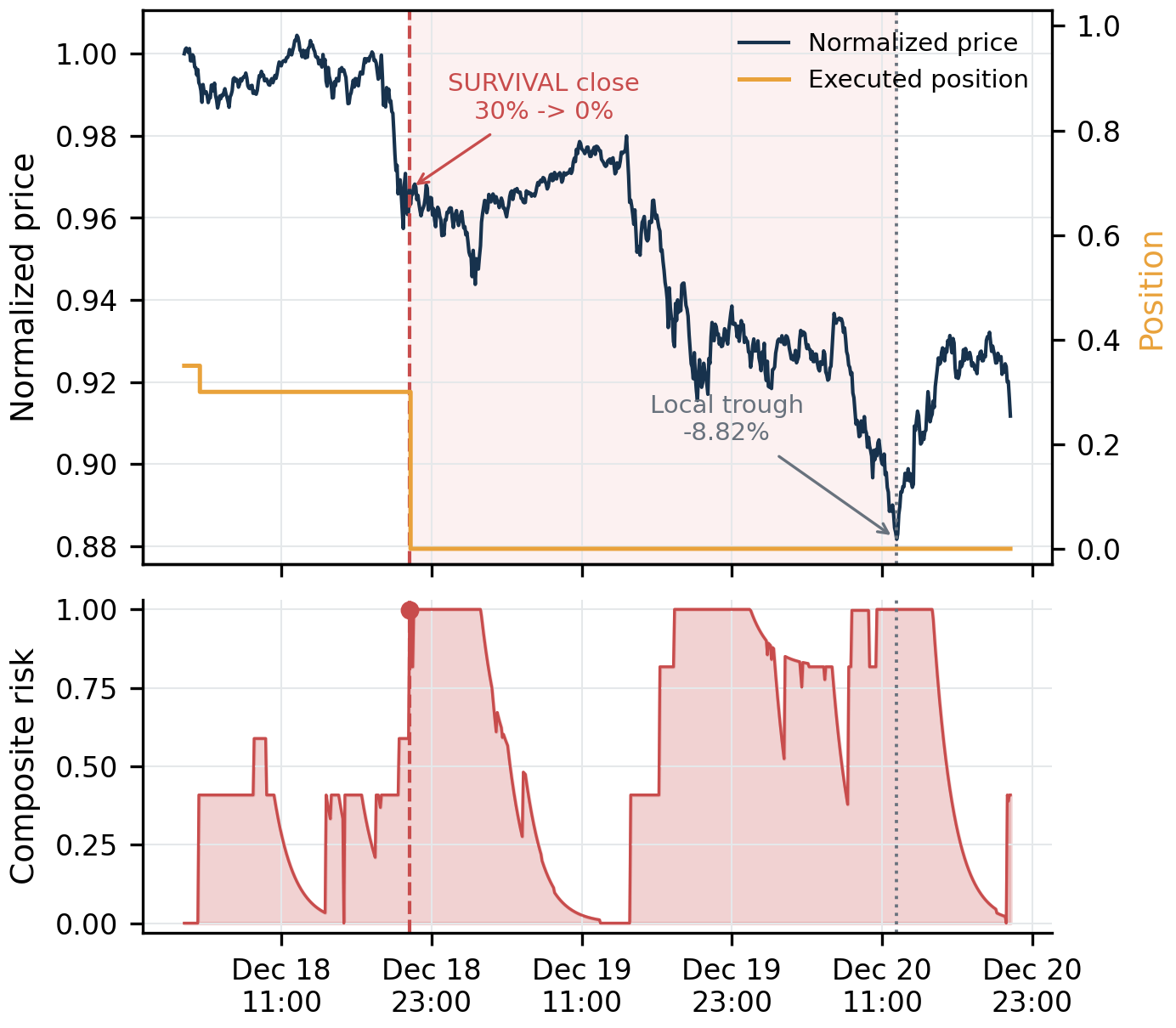}
\caption{BTC risk-to-execution trace around the December 18 \textsc{survival}
intervention. The controller closed a 30\% position before an 8.82\% decline
to the subsequent local trough.}
\label{fig:risk_position_case}
\end{figure}

\begin{figure*}[t]
\centering
\includegraphics[width=0.90\textwidth]{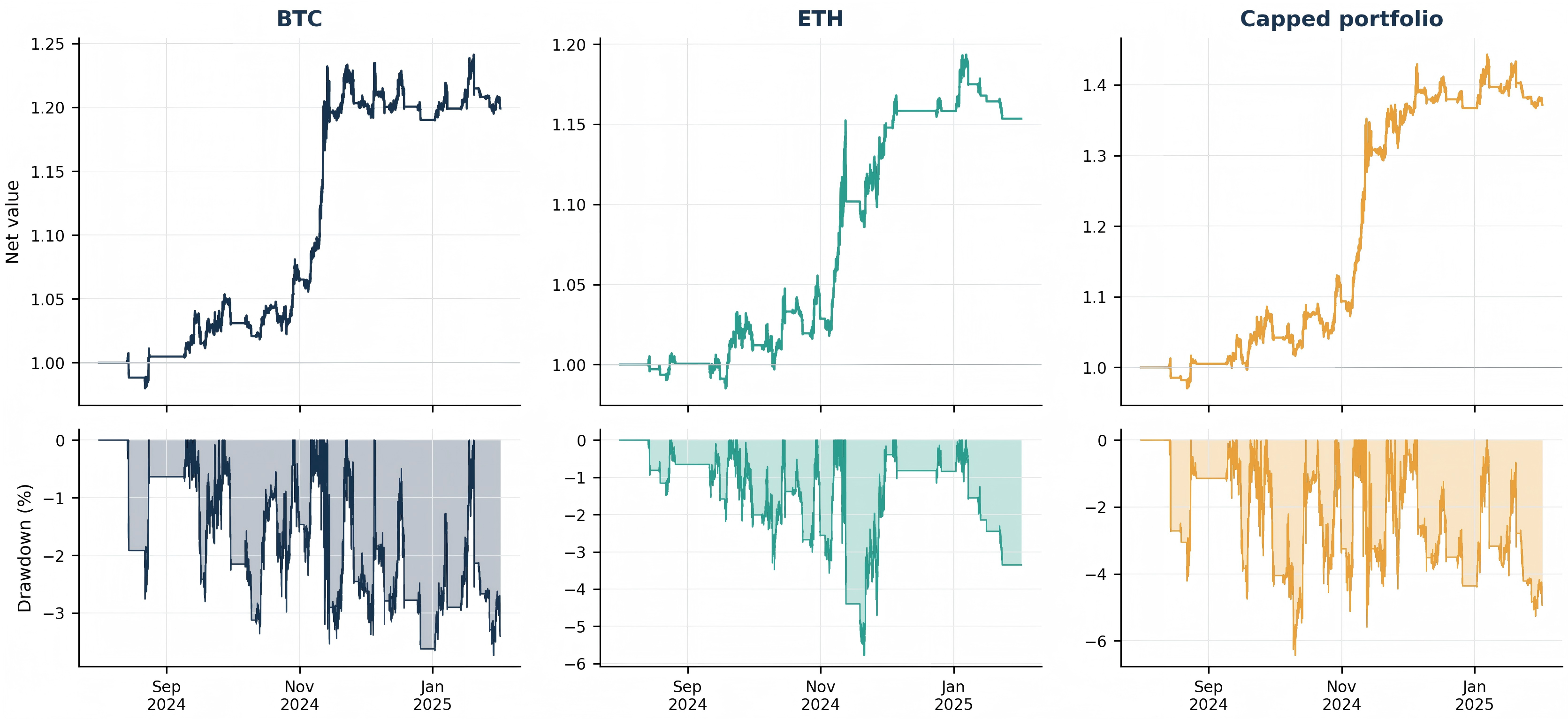}
\caption{Cost-adjusted five-minute net value and drawdown paths for SwingGuard on
BTC, ETH, and the exposure-capped joint portfolio.}
\label{fig:equity_drawdown}
\end{figure*}

\subsection{Risk-to-Execution Auditability}

Figure~\ref{fig:risk_position_case} traces a representative intervention
while the system held an active BTC position. At 21:15 on December 18,
the controller entered the \textsc{survival} response and closed the 30\%
exposure. BTC subsequently declined 8.82\% to the local trough on December
20 while executed exposure remained zero. The trace links detected risk,
state transition, controller action, and realized position over the subsequent
decline.

Figure~\ref{fig:equity_drawdown} shows shallow drawdowns throughout the
five-minute paths. The BTC and ETH equity curves
rise most during the shared November opportunity and then consolidate without
surrendering the accumulated gain. The December trace complements this aggregate pattern:
nonzero exposure precedes escalation, the state changes before the trough,
and exposure remains zero during the decline. Together they connect state
transitions to trading actions; the controlled comparison below then tests
the incremental effect of pre-generation conditioning.

\subsection{Controlled Ablation Results}

Table~\ref{tab:ablation_results} reports the configurations in
Table~\ref{tab:ablation_matrix} on the exposure-capped BTC--ETH portfolio.
Full versus Post-hoc Risk tests when risk enters strategy generation; Post-hoc
Risk versus No Risk Controller tests downstream governance; the
final two variants measure the Analyst Team and debate components.

\begin{table}[!htbp]
\centering
\small
\setlength{\tabcolsep}{3pt}
\begin{tabular*}{\columnwidth}{@{\extracolsep{\fill}}lccc@{}}
\toprule
Mode & Sharpe $\uparrow$ & Return (\%) $\uparrow$ & MDD (\%) $\uparrow$ \\
\midrule
Full SwingGuard & \cellcolor{metricblue!60}\textbf{2.98} & \cellcolor{metricorange!60}\textbf{37.16} & \cellcolor{metricgreen!60}\textbf{-6.44} \\
Post-hoc Risk & 2.79 & \cellcolor{metricorange!25}35.03 & -8.62 \\
No Risk Controller & 1.87 & 23.02 & -14.19 \\
No Analyst Team & \cellcolor{metricblue!40}2.88 & \cellcolor{metricorange!40}35.96 & \cellcolor{metricgreen!40}-6.71 \\
No Bull/Bear Debate & \cellcolor{metricblue!25}2.82 & 34.90 & \cellcolor{metricgreen!25}-7.03 \\
\bottomrule
\end{tabular*}
\caption{Controlled ablation results for the exposure-capped BTC--ETH
portfolio. Blue, orange, and green identify the top three results for Sharpe,
return, and MDD, respectively; the best result is boldfaced. MDD is reported
as a negative return, so higher is better.}
\label{tab:ablation_results}
\end{table}

Full SwingGuard ranks first on all three metrics. Against Post-hoc Risk, it
increases Sharpe from 2.79 to 2.98 and return from
35.03\% to 37.16\%, while improving MDD from -8.62\% to -6.44\%. Because these
configurations share evidence, strategy machinery, audit, and controller, the
observed difference is associated with placing risk before strategy
finalization. Sharpe and return improve by 6.8\% and 6.1\%, while drawdown
magnitude falls by 25.3\%, indicating that conditioning has its strongest
relative effect on downside quality.

With pre-generation conditioning absent, removing downstream governance
produces the largest adjacent deterioration:
Sharpe falls from 2.79 to 1.87, return from 35.03\% to 23.02\%, and MDD
deteriorates from -8.62\% to -14.19\%. Removing the Analyst Team or debate yields smaller but
directionally consistent losses:
Sharpe declines to 2.88 and 2.82, respectively, with lower returns and higher
drawdowns. This ordering associates the largest increment with five-minute
governance, a further improvement to upstream conditioning, and complementary
gains to specialization and deliberation. Together, these
comparisons separate two sources of stability: downstream governance limits
realized exposure between revisions, while pre-generation conditioning changes
the proposal entering that layer. Their combined advantage reflects coordinated
intervention timing and proposal formation.

\FloatBarrier

\section{Conclusion and Future Work}

SwingGuard formulates cryptocurrency swing trading as dual-timescale control:
daily multi-agent reasoning forms a risk-conditioned target, while the
five-minute Risk Controller maintains the System Risk Context, constrains exposure, and records
the path from evidence to execution. This coordination lets risk shape proposal
formation and maintain exposure compatibility between daily revisions.

Across BTC and ETH, SwingGuard remains profitable with the highest Sharpe and
lowest drawdown point estimates among reported methods. The capped portfolio
also earns a higher return than equal-weight buy-and-hold at lower exposure.
Monthly results
support cross-asset consistency, while the event trace establishes auditable
state-to-position linkage.

Ablations support both risk pathways, with the largest deterioration after
removing downstream governance. Supporting ablations distinguish this primary
risk-governance contribution from the complementary effects of specialized
analysis and deliberation.

Future work will test broader regimes and recovery efficiency.

\FloatBarrier

\section*{Acknowledgments}

This work was supported in part by Ripple under University Blockchain Research
Initiative (UBRI)~\cite{rippleubri}.

% References generated from the verified BibTeX database with aaai2027.bst.
\bibliography{swingguard_references}

\end{document}